\documentclass[pra, a4paper, twocolumn]{revtex4}   
\usepackage{graphicx}

\begin{document}
\title{Cs CPT magnetometer  for cardio-signal detection in unshielded
  environment} 

\author{J.\ Belfi$^{1}$, G.\ Bevilacqua$^{2}$, V.\
  Biancalana$^{1,*}$, S.\ Cartaleva $^{3}$, Y.\ Dancheva$^{1}$, L.~Moi$^{1}$ }

\address{$^1$CNISM-Unit\`a di Siena, Dipartimento di Fisica \\
Universit\`a di Siena, via Roma  56, 53100 Siena, Italy}

\address{$^2$CNISM-Unit\`a di Siena, Dipartimento di Ingegneria dell'Informazione \\
Universit\`a di Siena, via Roma  56, 53100 Siena, Italy}

\address{$^3$Institute of Electronics BAS, 72 Tzarigradsko Chausse,
  1784, Sofia, Bulgaria}

\address{$^*$Corresponding author: biancalana@unisi.it}

\begin{abstract}
We present first, encouraging results obtained with an experimental
apparatus based on Coherent Population Trapping and aimed at detecting
biological (cardiac) magnetic field  in magnetically compensated, but
unshielded volume. The work includes magnetic-field and
magnetic-field-gradient compensation and uses differential detection
for cancellation of (common mode) magnetic noise. Synchronous data
acquisition with a reference (electro-cardiographic or
pulse-oximetric) signal allows for improving the S/N in an off-line
averaging. The set-up has the relevant advantages of working at room
temperature with a small-size head, and of allowing for fast
adjustments of the dc bias magnetic field, which results in making the
sensor suitable for detecting the bio-magnetic signal at any
orientation with respect to the heart axis and in any position around
the patient chest, which is not the case with other kinds of magnetometers.
\end{abstract}


\maketitle

\section{Introduction}
Spectroscopy of  transitions involving long-lived levels in alkali
atoms has shown  impressive potentialities  in the field of high
resolution and high sensitivity magnetometry since the birth of
coherent spectroscopy \cite{{Blo62},{Cohen69},{Bud06}}. In the last years, many research groups, with different
experimental techniques like non linear magneto-optical
rotation\cite{budk_rev02}, spin-exchange relaxation free - Faraday rotation \cite{romalisSERF}, double resonance optical
pumping \cite{Groe06}, and Coherent Population Trapping (CPT)
\cite{Kna02,NOI} have obtained important records in optical
magnetometry  sensitivity, so to prove that atomic magnetometers are
presently competitive  even with SQUID magnetometers in terms of
sensitivity, and can find applications in fields like geomagnetism
monitoring, testing of materials, testing of fundamental symmetries of
Physics and bio-magnetism detection.   

The first mapping of the cardiomagnetic field with a Rubidium double
resonance magnetometer has been demonstrated in ref.\cite{wei03a}, and
good agreement with  the magnetocardiogram obtained by a SQUID
magnetocardiograph has been found\cite{wei03b}. 

The unique experiment of M. Romalis and coworkers \cite{Xia06},
recently showed that laser magnetometers can perform records in
registration of  such weak (few hundreds fT) magnetic fields as those
produced by the human brain activity. 

A fundamental request for  actually reaching the  sensitivity level
needed for bio-signal detection is to reduce dramatically the
environmental magnetic field. Usually high magnetic permeability
materials like  $\mu$-metal are employed  to build small volumes
inside of which  it is possible to have very effective  shielding of
the external, undesired magnetic fields. 

The high cost  of such magnetically clean chambers represents a strong
limitation for the diffusion  of high sensitive magnetometry in
clinical application, also because of their delicateness and
possible magnetization with consequent requirement of periodical
demagnetizing treatments.   
 
The main limits of working in unshielded environmental conditions,
typical in a scientific laboratory, are given by the presence of
strong magnetic field gradients, ac magnetic fields in the volume of
sensor cell and by the magnetic noise produced by other human
activities and ionospheric phenomena.  

Field gradients  affect the sensitivity limit of the instrument by
introducing  a broadening of the detected resonance
line. Time-dependent magnetic fields represent a further  additional
source limiting the magnetic field detection sensitivity.  

The possibility of achieving high sensitivity, simply by passively
quenching magnetic noise (placing  thick Al plates around the magnetic
sensor), has been shown in ref.\cite{seltz04}. There a sensitivity  of
the order of $1\rm{pT}/\sqrt{\rm{Hz}}$ is obtained thanks to a
sophisticated feedback system that, keeping around  zero the total
magnetic field value in the volume of a K  cell (heated to
$170^{\circ}C$), suppresses the broadening due to
spin-exchange-collisions.    

In the present work, we show that a cheap system for eliminating
spatial inhomogeneities of the background magnetic field, joined to an
accurately balanced differential detection scheme, allows for
detecting the magnetic signal produced by the heart activity
without the use of neither expensive and bulky shielding chambers nor
aluminum shields that attenuate high-frequency magnetic noise. The
device consists in a full optical sensor operating at room
temperature, placed inside a frame of compensation coils that can be
easily accessed by a human body.

\section{Basics of the measurement principle}
\label{principles}
Magnetic field optical detection is performed by creating
CPT\cite{NOI, And03} on   Zeeman sublevels of the F=3 hyperfine
ground state of Cs atoms.  

This is an all optical technique as it does not require the presence
of coils in the proximity of the sensor cell in order to produce
direct RF magnetic excitation. This feature makes easy to optimize the
sensitivity to any orientation of the magnetic field, thus allowing
for measuring any component of it. CPT is produced  when circularly polarized
laser radiation, resonant with a $\rm{D}_2$ optical transition, is frequency
modulated exactly at the Zeeman frequency splitting of the ground
Zeeman sublevels. The CPT signal can be seen as a resonant decrement
of the absorption in the optical transition. The measure of the
resonant frequency $\omega_L$ gives directly an absolute measurement
of the modulus of the magnetic field interacting with the atomic probe.  

 This kind of magnetic field measurement is absolute and
 self-calibrated because the magnetic field strength $B$  is given by: 
$B = \frac{\omega_L}{\mu_0 g_F}$, where  $\mu_0$ is the Bohr magneton,
 $g_F$ is the Land\'e factor of the considered ground-state and
 $\omega_L$ is the resonance  (Larmor) frequency. The magnetometer is
 a scalar one and the  estimation  of the  magnetic field modulus is
 obtained  by scaling  the Larmor frequency by atomic constants, and
 only minor deviations  may occur \cite{Wyn99}, introducing small
 systematic errors. Long time  living coherence between Zeeman
 sublevels allows for creating  very  narrow (few Hz) resonances and
 then for getting  very sensitive  magnetic field estimations.  Cs
 optical transitions can be excited by means of diode laser
 radiation, that can be easily  modulated in frequency  by means of
 the modulation of its injection  current. Magnetic field intensities in the
 geophysical range, correspond to  frequency modulations of the order
 of 100~kHz. This  gives the  possibility of performing high precision
 magnetic field measurement while  operating in  the presence of the
 local Earth  magnetic field, by  employing  RF modulation of the
 laser current\cite{budk_mod_02,Aco06, NOI}. 

The presented experimental investigation is dedicated to the
detection of a small, time varying  field in the presence of about
$10^6$ times stronger dc bias field, for any  relative  orientation
of the two fields. As any scalar magnetometer, our setup is sensitive to the component of the
varying magnetic field in the direction of the bias field (see also Section
\ref{inhomogeneities}).

\section{Sensor set-up}
\begin{figure}[h]
\begin{center}
\includegraphics[width=7cm]{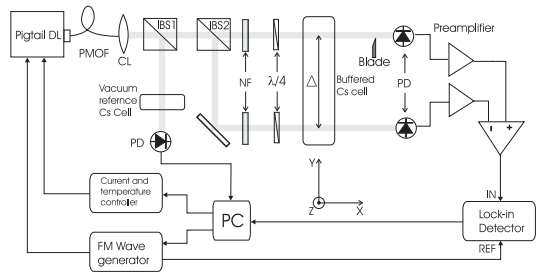}
\end{center}
\caption{ Experimental setup. PMOF: polarization maintaining optical
  fiber, CL:collimating lens, IBS: intensity beam splitter, 
    NF: neutral filters, PD: photodiode. }\label{setupSINGLEcell}
\end{figure}

The sensor set-up is sketched in Fig.\ref{setupSINGLEcell}.
Laser source is a single-mode edge-emitting pigtail laser ($\lambda$=
852~nm) with 15~mW of laser power and an intrinsic line-width of
less than 5~MHz. Laser light is coupled to a single-mode
polarization-maintaining fiber 10~m in length and the laser head,
containing the laser chip, a 40~dB optical isolator, the temperature
stability system (Peltier junction and the thermistor), and the fiber 
collimator is contained into a compact butterfly housing of  $\mbox
40~\textrm{cm}^3$.  

The beam exiting from the fiber is  collimated  then  split by an
intensity beam splitter (IBS1). The beam reflected by IBS1 passes
trough a vacuum reference cell and the transmitted light is detected
in order to monitor the laser frequency tuning with respect to the
Doppler line. The beam transmitted trough IBS1 is split again by a
second intensity beam splitter IBS2. The  beam reflected by IBS2 is
made parallel  to the transmitted one after  reflection on a mirror M
and plays the role of a second arm of the sensor. The two arms
interaxial separation $\Delta$ is chosen to be about 7 cm.  The
polarization of both the beams is adjusted by means of two $\lambda/4$
plates. Both beams cross, in different points,  a single cell,
containing Cs and 5~Torr of $Ar$ as a buffer gas.  The use of a single
cylindrical cell, 9~cm in length and 1.5~cm in radius guarantees equal
conditions for CPT creation (equal Cs and Ar vapor density in the two
laser-atoms interaction volumes). All the component represented 
in Fig.\ref{setupSINGLEcell} are mounted on a non magnetic plate and the 
overall dimensions of the sensor are $25\times12\times4$~$\rm{cm}^3$. 

In order to reduce power broadening, laser power  is reduced to less than
$50~\mbox{$\mu W$}$ in each sensor arm using a set of neutral
filters. A non magnetic translating  blade is placed after the
cell, in front of one (the most sensitive) of the two photo-diodes, in
order to have the possibility of fine balancing the detected
photo-currents.

In order to increase the signal to noise ratio of the detected signal,
phase sensitive detection is used. At this aim, we impose a 20~kHz
frequency modulation on the RF frequency modulating the laser current
and then detect the component of the signal oscillating at this frequency.  

\section{Sensitivity optimization for cardio-signal detection}
The magnetic field produced by a  human heart is characterized by a
maximum intensity of about 100~pT in the close vicinity of the chest,
and the main features are well reconstructed if sampled in a bandwidth
of at least 30~Hz \cite{wei03a}.  

Magnetic heart-beat spatially confined distribution, of the order of the
heart dimensions, allows for performing differential detection in a
rather small volume in front of the patient's chest, while the
possibility of triggering the acquisition on other signal, taken for
example from the electro-cardiogram (or even from a simple pulse
oximeter), allows for effective noise rejection in an off-line analysis. 

As introduced before, unshielded environmental conditions limit the
instrumental sensitivity essentially because of  the temporal and 
spatial fluctuation of the background magnetic field. Field gradients
in our laboratory have the typical magnitude of about 100~$\rm{nT}/\rm{cm}$
depending on the presence of magnetic material both in the
instrumentation and in the building structure. Alternating magnetic
fields contribute instead, mainly with $50~\rm{Hz}$ and higher odd
order harmonics, with a typical rms amplitude of the order of 100~nT
depending on the vicinity of electric lines and power supply
transformers. For  frequencies higher than the lock-in cut-off
frequency, ac fields determine  an effective broadening of the CPT
line, while  for lower frequencies they directly introduce time
fluctuation  of  the CPT resonance center. Optimization of the
magnetometer sensitivity is fundamentally performed on the basis of
the spatial and temporal characteristics of the bio-signal to be measured.  

In order to improve the common mode noise rejection it is important to
have, in absence of the magnetic source to be measured, exactly the
same resonance profile at the  two inputs of the differential lock-in
amplifier. As shown in Ref.\cite{Kna02} small differences in  laser
power, polarization, beam-size contribute significantly to the
unbalance  by changing the amplitude and the shape (mainly) of the
CPT resonance. Difference between the magnetic field values inside the
two laser-atom interaction volumes shifts instead the positions of the
resonances centers.  

The cardiac signal measurement procedure consists in, first, 
adjustment of  amplitude and width of the CPT signals in the two arms
separately (single input mode of the lock-in amplifier) and, second,
fine optimization of those parameters by looking at the maximum
common mode noise cancellation in the differential signal
(differential input mode of the lock-in amplifier).

\subsection{Field inhomogeneities compensation}
\label{inhomogeneities}
As mentioned in Section \ref{principles}, measuring the modulus of $B$
in the presence of a strong dc component, leads to detect only the
variations $\Delta B$ of the field in the component parallel to the dc, bias
field since 
$\Delta \left|\vec{B}\right|\simeq
\frac{\vec{B}}{\left|\vec{B}\right|}\cdot\Delta\vec{B}$. 

By zeroing two of the three spatial component of the local background
magnetic field, one is  able to measure one selected  component of the
signal produced by the biological source. This gives the possibility
of studying   the different components of the measured field and
eventually  reconstructing the entire vectorial signal. 

\begin{figure}[h]
\begin{center}
\includegraphics[width=6cm]{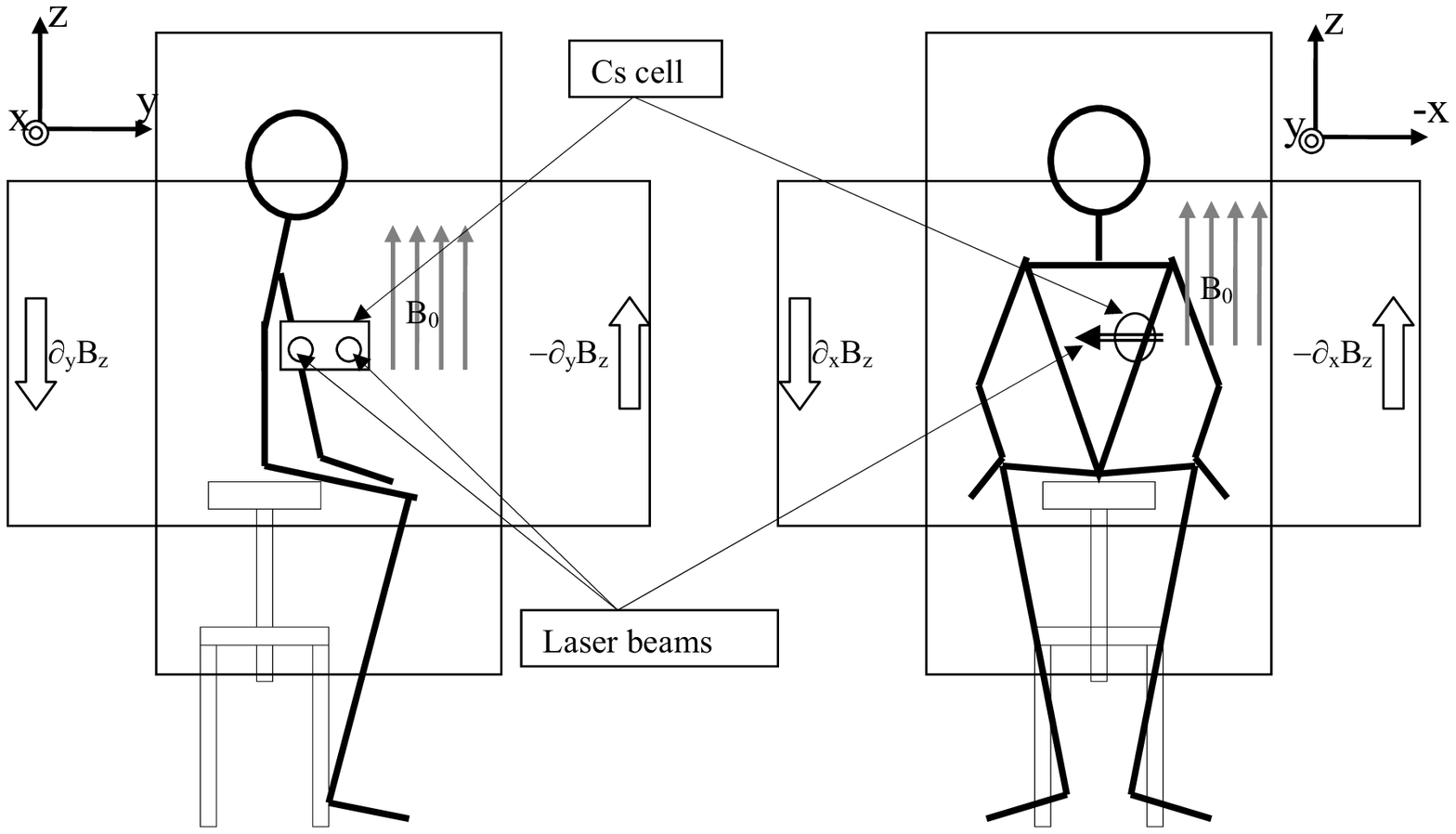
}
\end{center}
\caption{Schematics of the apparatus.  The selected component of the background
bias field  $B_0$, parallel to $\hat{z}$ in the scheme, determines the
direction of the detected component of the bio-signal, while laser
beams separated by  $\Delta$ are directed along $\hat{x}$. Dipoles are used
for gradients compensation.  \label{cubo_paziente}} 
\end{figure}

The typical configuration for  a cardio-signal detection measurement
is sketched in Fig.\ref{cubo_paziente}. In this particular
configuration both $x$ and $y$ components of the laboratory
magnetic field are compensated (the residual magnetic field is in the
0.1~${\rm{\mu} T}$ range) and thus, the $z$ component of the human 
heart magnetic beat is being detected.

This  detection scheme has important advantages in view of simplifying the compensation of 
magnetic field inhomogeneities. As explained in deeper detail in the Appendix, 
in fact, only 3 currents for the compensation of the spatial gradients inside the cell volume,
are needed in addiction to the 2 currents employed for the zeroing of
two of the three component of the laboratory magnetic field. These three
currents  flow respectively in one anti-Helmholtz pair  and in two
small coils playing the role of magnetic dipoles. 

\begin{figure}[h]
\begin{center}
\includegraphics[width=7cm]{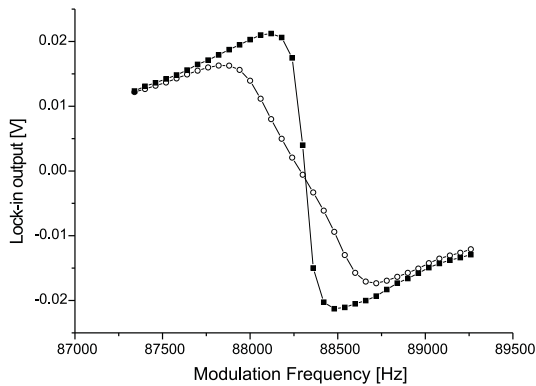}
\end{center}
\caption{CPT line-shapes produced with circular polarization and
  magnetic field orientation orthogonal to the laser beam (along
  $\widehat{z}$ direction). Circles: static gradients are not
  compensated. Fitted line-width is 1.67~kHz. Squares: optimized
  compensation of the gradient components. Fitted line-width is
  reduced to 280~Hz.  \label{restringimento}} 
\end{figure}

After the optimization of the gradient compensation we get a strong
reduction of the line broadening due to magnetic field
inhomogeneities. Fig. \ref{restringimento}  shows the CPT resonance
signal registered with and without the compensation of spatial
gradients. Without compensation a best fit analysis gives a 
linewidth (FWHM) of  about 1.7~kHz which is reduced down to less than
300~Hz in case of gradients compensation.  

\subsection{Differential measurement procedure and noise characterization}

Depending on the spatial configuration of the local magnetic field in
correspondence of  the two light-atoms interaction volumes, we are
able to cancel different noise contributions at the cost of increasing
the uncorrelated electronic noise in the detection stage by namely a
factor of  $\sqrt{2}$. Other noise sources, like laser frequency noise
and amplitude noise are systematically canceled from the signal
provided that the two arms are  balanced. If, furthermore, both arms
of the sensor give CPT resonance exactly at the same resonant
frequency, then even the common mode magnetic field noise, generated
for example by all far magnetic sources is eliminated. 

\begin{figure}[h]
\begin{center}
\includegraphics[width=7cm]{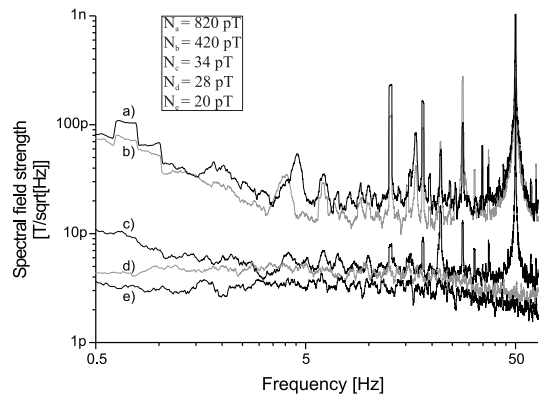}
\end{center}
\caption{a)~Single input acquisition: laser frequency is tuned to the
  center of the CPT resonance. b)~Differential input acquisition: CPT
  signals  on the two arms of the sensor are separated by more than
  the CPT linewidth. c)~Differential input acquisition: CPT signals
  overlapping is optimized. d)~Out of Doppler absorption resonance
  noise in differential input mode. e)~Out of Doppler absorption
  resonance noise in single input mode. In the inset is reported  the total rms noise $N$ integrated betweeen 1 Hz and 30 Hz (cardio-signal bandwidth) relative to each trace.  \label{cardiac_noise}} 
\end{figure}

In Fig.~\ref{cardiac_noise} is presented the magnetometric noise spectral
density for different operating conditions of the device. Trace (a) is
relative to a single arm operation: laser noise and magnetic common mode
noise contribute entirely. Trace (b) is obtained in differential input
acquisition, where the two CPT resonances occur at frequencies
separated by more than the CPT linewidth. In this case one mainly gets
the cancellation of the laser frequency and intensity noise. Evident
gain in the reduction of residual noise is obtained in the
differential acquisition when CPT resonances detected by the two
sensors are accurately overlapped (trace (c)).  Trace (d) and (e) are
recorded respectively in differential and single input mode, when the
laser frequency is tuned  out of the single photon (Doppler) resonance. It
is evident that the uncorrelated electronic noise increases in the
differential detection.

\section{Cardio signal data analysis}
The signal-to-noise (S/N) ratio is not high enough to make single
cardiac pulse directly observable, so that off-line analysis must
include both linear filtration  and averaging of the data. We use a standard
technique for acquiring a  reference signal simultaneously with the
magnetometric data. Our digital  lock-in amplifier (Stanford SR830)
allows for storing up to 16384 data in a  buffer, which can be
synchronously acquired using an external clock.  Thus we use an
external DAQ card (MCC 1608FS, 16~bit resolution, with USB  interface)
whose ADC operation triggers the lock-in data storage. Such DAQ is
used to sample reference data produced by either the pulse-oximeter or
an  electrocardiographic connection.  

In the case of the pulse-oximeter, we arranged an IR led and a 
phototransistor in such a way  to sense the transparency 
variation of a finger, and give several hundred mV signal
peak-to-peak. Alternatively, the ECG was performed by acquiring digitally the
electric signal collected by two electrodes placed in the vicinity of
the heart, after a passive low-pass filtering, necessary to reduce the
noise aliasing. Specifically, as we use typically a 128~S/s
acquisition rate, a 18~dB/oct filter made with a three-stage RC
cutting at 40~Hz, produced a clean signal about 1~mV at the  
QRS complex.

In both cases, the reference signal needs some preliminary (off-line) 
numerical conditioning before being used as a trigger. Namely, some linear 
filtering is performed to remove slow drifts and high frequency noise (a 
3rd order bandpass numerical filter from 0.2~Hz to 30~Hz is usually 
suitable). After this, the envelope of the reference is evaluated, and used 
to normalize the peak-to-peak amplitude all along the registration. 
Finally, the normalized reference is used to produce a trigger signal 
which is adjustable in slope, threshold and hold-off, similarly to standard 
oscilloscopes. Additionally, and in particular when using pulse-oximeter 
reference, it is important to adjust a (negative) trigger delay, of
some tenths of second, due to the  relevant lag between the heart
pulse and the change on the finger  transparency. Both references
showed to be suitable for data average triggering. 

The magnetocardiographic data downloaded from the lock-in data buffer are 
firstly numerically filtered with the aim of removing specific spectral 
components (such as 50~Hz noise) and other spectral peaks due to artifacts (such as mechanical vibration of the Helmholtz coils) then split in traces 
corresponding to single heart pulses, from which the average pulse can be 
reconstructed by  averaging.

The averaging process is not straightforward, due to the fact that the 
pulse duration is not stable. We tried two different approaches to superpose and average 
the single-pulse traces. In the first one, we strengthen the arrays 
containing the single pulse traces to a given size by linear
interpolation, to make all  of them equal in length; while in the
second procedure we add zeros at  the end of the shorter arrays to
make all of them as long as the longest  one. It turns out that the
second procedure is more effective and correct,  due to the fact that
(as can be seen also by the ECG traces) the traces  corresponding to
single pulse cardiac activity are not similar to each other. 

Specifically, each trace starts with a P-QRS-T complex having a pretty 
stable duration, followed by a quiescent interval whose duration is rather 
variable in time, which is the main responsible for the
non-periodicity of the signal.  As a consequence, using ECG
reference, one finds that the averaging is the best (and the best
is the S/N obtained) when triggering on the QRS complex (the highest peak)
with a delay suitable to include the P wave of the same beat,
while the location where the quiescent phase is being cut is not
relevant, provided that the T wave of the previous beat is not
included. In the case of pulse-oximeter as a reference the correct
trigger delay must be found empirically due to its intrinsic additional delay.

\begin{figure}[h]
\begin{center}
\includegraphics[width=7cm]{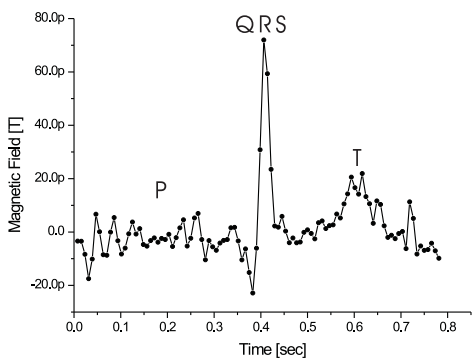}
\end{center}
\caption{Reconstructed cardiac pulse. The signal is averaged over a
  set of about 150 cardiac pulses acquired at 128~Hz sampling rate. The
  lock-in equivalent noise band-width is 41.7~
  Hz.  \label{Single_cardiac_pulse}}   
\end{figure}

In Fig. \ref{Single_cardiac_pulse} is reported a reconstructed
magnetic heart-beat, obtained by placing the sensor close to the chest of
one of the authors. The peak is obtained from a data set consisting in
16384 points acquired at a sampling rate of 128~Hz. Average is
 thus performed on a set of about 150 heart-beats, corresponding to 128~sec
of measuring time. 

\section{Conclusions}
We have demonstrated the potential of a fully optical, CPT based, differential
magnetometers for magneto-cardiographic applications in unshielded
environment, using a sample of Cs vapor at room temperature in a compact
sensor head. We discussed simple approaches for
magnetic-field and magnetic-field-gradient compensation, as well as
off-line data analysis of quasi-periodic signals.
The noise characterization showed that a severe limitation to the
sensitivity would be set by the background magnetic field fluctuation,
which, thanks to its essential uniformity can be effectively canceled in
differential measurements. Specific features of the presented
magnetometer, compared to others used at present lie in the facts that
it works at room temperature and can be fast adjusted to detect
different components of the weak biologic magnetic field vector, as
the sensor, being purely optical, does not include coils,  which would
constitute geometrical constraints. The simplicity of the method, its low cost
and low maintenance cost can be a crucial factor for its
dissemination in clinical applications.

\appendix

\section*{Magnetic-field and gradient compensation}
In this appendix, we report the principles of the strategy  for the resonance
narrowing, shown in Fig.~\ref{restringimento}, obtained by means of
magnetic field and magnetic field gradient conditioning. 

In order to be able to independently compensate each of the three
components of the magnetic  field vector, while introducing minimum
inhomogeneities, it is straightforward to  use three pairs of coils in
Helmholtz configuration (which for squared  coils means to place them
at a distance of about 0.544~L, L being the length of the coil side).

The center  of the Cs vapor cell, the core of the optical sensor, is
then placed at the geometrical  center of  a "cube" composed by a set
of three pairs of squared ($180\times180\,\rm{cm}^2$) coils
($180\times0.544$ is just enough to make possible a human body to
access the cube).

  
Basic arguments  give that not only $\vec{\nabla}\cdot\vec{B}$, but
also $\vec{\nabla}\times\vec{B}$ is zero in the position of the
sensor, as no current flows there. As a  direct consequence, only five out of 
the nine elements of the matrix $\partial B_i/\partial x_j$ are 
independent, i.e. only five parameters have to be controlled for their 
compensation. More precisely, a complete compensation requires to controll 
two of the three diagonal elements, and three of the six 
off-diagonal elements.

The diagonal elements can be compensated by anti-Helmholtz coils. This 
means that two additional couples of coils with counter-propagating 
currents could be used to compensate $\partial B_x/\partial x $ 
and  $\partial B_y/\partial y $ (which would guarantee also the 
compensation of $\partial B_z/\partial z $). In principle, such 
anti-Helmholtz coils should be at a larger distance, in order to produce 
the linear term $\partial B_i/\partial x_i$ and not the cubic one. But for 
simplicity, considering the small corrections to be made, is preferable to 
use the already existing Helmholtz pairs (with which, by the way, one has 
also the advantage of producing a maximum gradient, given the current).

This choice leads to simultaneously compensate the $x$ and
$y$ components of the field and the three diagonal components of  the
gradient, by separately controlling the currents in the four Helmholtz
coils for  $B_x$ and $B_y$ compensation. Specifically, indicating with
$I_{x1}$ and $I_{x2}$ the currents in the coils around the $x$ axis, $B_x$ is
controlled by $(I_{x1}+I_{x2})$ and $\partial B_x/\partial x $ is
controlled by $(I_{x1}-I_{x2})$ (similarly for the y direction). If
only $B_x$ and $B_y$ are to be compensated, so that only $B_z$
inhomogeneities are relevant a simpler control can be made with
three currents only, using pure Helmholtz configuration in $x$ and $y$
directions ($(I_{x1}=I_{x2})$, $(I_{y1}=I_{y2})$), and pure anti-Helmholtz 
$(I_{z1}=-I_{z2})$ configuration in the $z$ direction. In our present 
set-up, we opted for this latter, simpler choice.

The three independent off-diagonal elements of the gradient ($\partial 
B_y/\partial x$, $\partial B_z/\partial x$, $\partial B_z/\partial y$) can 
be controlled with couples of magnetic dipoles symmetrically located far 
away from the sensor in order to produce a vanishing, quadrupole field in 
the region of the cell. Again, provided that $\partial B_y/\partial x$ 
produces negligible effects because not responsible for $B_z$ 
inhomogeneities, only two of these three couples needs to be
effectively  adjusted. We actually use dipoles oriented in $z$
direction and  located in the $xy$ plane. As the dipole  field gradient 
decreases with the fourth power of the distance, fixing the dipoles on the
frame of the large Helmholtz coils  would make necessary to use pretty
large current and pretty heavy coils. We  simplified our task by using
several couples of Nd permanent magnets in order  to coarsely compensate
the off-diagonal gradient components, and smaller  electromagnets with
few hundreds mA current for fine adjustments.

\acknowledgments
The authors thank R.\ Mariotti and  M.\ Focardi for very useful discussions. 
S. Cartaleva acknowledges CNISM for the grant (Code: FOES000020/ref. num. OA 06000089). 
This work was supported by the Monte dei Paschi di Siena Foundation.

\end{document}